\documentclass{aa}
\usepackage{graphicx}

\begin{document}

\title{Fourier decomposition and frequency analysis of the pulsating 
       stars with ${\rm P} < $ 1d in the OGLE database}
\subtitle{ II. Multiperiodic RR Lyrae variables in the Galactic Bulge}

\author{P. Moskalik\inst{1} \and  E. Poretti\inst{2}}

\institute{
Copernicus Astronomical Center, ul. Bartycka 18, 00-716 Warsaw, Poland 
\and
INAF-Osservatorio Astronomico di Brera, Via Bianchi 46, I-23807 Merate, Italy
}

\offprints{P. Moskalik,\\ \email{pam@camk.edu.pl}}

\date{Received 10 September 2002 / Accepted 30 October 2002}

\abstract{
We present the results of a systematic search for multiperiodic pulsators
among the Galactic Bulge RR~Lyrae stars of the OGLE-1 sample. We identify
one "canonical" double-mode variable (RRd star) pulsating in two radial
modes. In 38 stars we detect secondary periodicities very close to the
primary pulsation frequency. This type of multiperiodic variables
constitute $\sim\! 23$\% of RRab and $\sim\! 5$\% of RRc population of the
Bulge.  With the observed period ratios of $0.95-1.02$ the secondary
periods must correspond to nonradial modes of oscillation.  Their beating
with the primary (radial) pulsation leads to a long-term amplitude and
phase modulation, known as the Blazhko effect.  The Blazhko RRab variables
occur more frequently in the Galactic Bulge than in the LMC. The opposite
tendency is seen in case of the RRd stars. The differences of incidence
rates are most likely caused by different metallicity of the two
populations.  We discuss pulsation properties of the OGLE-1 Blazhko stars
and compare them with predictions of theoretical models. We argue, that the
oblique magnetic pulsator model of Shibahashi (\cite{shi00}) cannot account for the
observations and should be ruled out.


}

\titlerunning{Multiperiodic RR~Lyrae variables in the Galactic Bulge}
\authorrunning{Moskalik \& Poretti}
\maketitle

\section{Introduction}

The Optical Gravitational Lensing Experiment (OGLE; 
Udalski et~al. \cite{uda92}) is
devoted to search for dark matter in our galaxy through the detection of
microlensing events. As a by-product of this program a large amount of
photometric data has been collected for stars in the Galactic Bulge and
Magellanic Clouds. In this series of papers we examine short period
pulsating variables discovered in the OGLE-1 Galactic Bulge sample (Udalski
et~al. \cite{uda94}, \cite{uda95}, \cite{uda96}, \cite{uda97}).  
Paper~I (Poretti \cite{por01}) was devoted to 
\emph{monoperiodic} $\delta$~Scuti and RR~Lyrae variables.  In the current
paper, second in the series, we will discuss \emph{multiperiodic} RR~Lyrae
stars.

The existence of multimode RR~Lyrae variables has been known since the
discovery of double-mode pulsations in \object{AQ~Leonis} 
(Jerzykiewicz \& Wenzel \cite{jer77}). 
In following years about 300 similar variables have been identified
in several stellar systems, including 181 stars in the Large Magellanic
Cloud (see Kov\'acs \cite{kov01} for comprehensive review).  All these variables
share common properties: they pulsate in \emph{two radial modes}, namely
fundamental and the first overone, with the period ratio of 
${\rm P}_{\! 1}/{\rm P}_{\! 0}=0.742-0.747$. These stars are commonly
referred to as RRd variables. 

In recent years a new class of multiperiodic RR~Lyrae variables has been
identified (Olech et~al. \cite{ole99}).  They are characterised by presence of
two or more very closely spaced frequencies, with period ratios in the
range of $0.95-1.05$. Such period ratios are incompatible with the radial
mode oscillations and point strongly towards presence of \emph{nonradial}
modes.  This new form of multiperiodicity was first discovered in the RRc
variables, but later was shown to occur in the RRab variables as well
(Moskalik \cite{mos00}). For most complete summary of recent observations see
Kov\'acs (\cite{kov02}).

Massive photometry collected during microlensing surveys is particularly
well suited for systematic search of multiperiodic RR~Lyrae stars.  Such a
search has already been performed in the Large Magellanic Cloud, using
MACHO data (Alcock et~al. \cite{alc00}; Welch et~al. \cite{wel02}).  In this paper we
present a complete inventory of multiperiodic RR~Lyrae stars in the OGLE-1
Galactic Bulge sample. Our primary goal is to establish the incidence rates
of different types of multimode oscillations in the Galactic Bulge
population.  Comparison between the rates in the Bulge and in the LMC can
shed new light on the physical conditions favouring such oscillations. We
also discuss pulsation properties of the identified multiperiodic RR~Lyrae
stars. Preliminary results of this work has already been published by
Moskalik \& Poretti (\cite{mos02}). Here we present the complete discussion of our
findings.

\section{Search for multiperiodicity}

The OGLE-1 database contains 215 RR~Lyrae stars, 150 of them have been
classified as RRab (fundamental-mode pulsators) and 65 as RRc (first
overtone pulsators). The photometric data span $\sim 900$~days, with
typically 130-150 I-band measurements published for each star. As a first
step, the lightcurve is fitted with the Fourier sum of the form

$${\rm m_I}(t) =  \langle {\rm m_I}\rangle + \sum_{k} {\rm A}_k \cos (2\pi k{\rm f}_0 t + \phi_k)\eqno(1)$$

\noindent The pulsation frequency ${\rm f}_0 = 1/{\rm P}_{\! 0}$ is also
adjusted in the fitting process.  For several stars slow instrumental
drifts are present in the data.  These are modeled by a cosine term with
period of 50000\,day added to Eq.\,(1).

In the next step, the search for additional periodicities is performed.
This is done with two different, independently applied methods:

\begin{enumerate}
\item The fit Eq.\,(1) is subtracted from the data. The Fourier power
      spectrum of the residuals is then computed in order to reveal 
      a secondary frequency, if present.

\item The fitting formula Eq.\,(1) is supplemented with an additional 
      cosine term with frequency ${\rm f}_1$. The data is then fitted for 
      different trial values of ${\rm f}_1$, keeping fixed the primary 
      frequency ${\rm f}_0$ and the number of its harmonics, but 
      recalculating their amplitudes and phases for each trial. A secondary
      frequency, if present in the data, should reduce the dispersion of the 
      fit in a significant way.
\end{enumerate}

\noindent The two procedures give the same results for all the stars except 
one, strengthening our confidence in the frequency identifications. In the
following, we accept as multiperiodic only those variables in which
secondary period is detected with both methods.

As a third step, the Fourier fit with two identified frequencies and 
\emph{their linear combinations} is performed.  To this effect, we fit the
lightcurve with the following formula:

$${\rm m_I}(t) = \langle {\rm m_I}\rangle + \sum_{k,n} {\rm A}_{kn} \cos [2\pi (k{\rm f}_0+n{\rm f}_1) t + \phi_{kn}]\eqno(2)$$

\noindent The search for additional periodicities is then repeated, again using both
methods: Fourier transform of residuals (method 1) and minimization of
dispersion with respect to a new trial frequency ${\rm f}_2$ (method 2). The
process is stopped when no significant new terms appear.

\begin{figure}
\resizebox{\hsize}{!}{\includegraphics{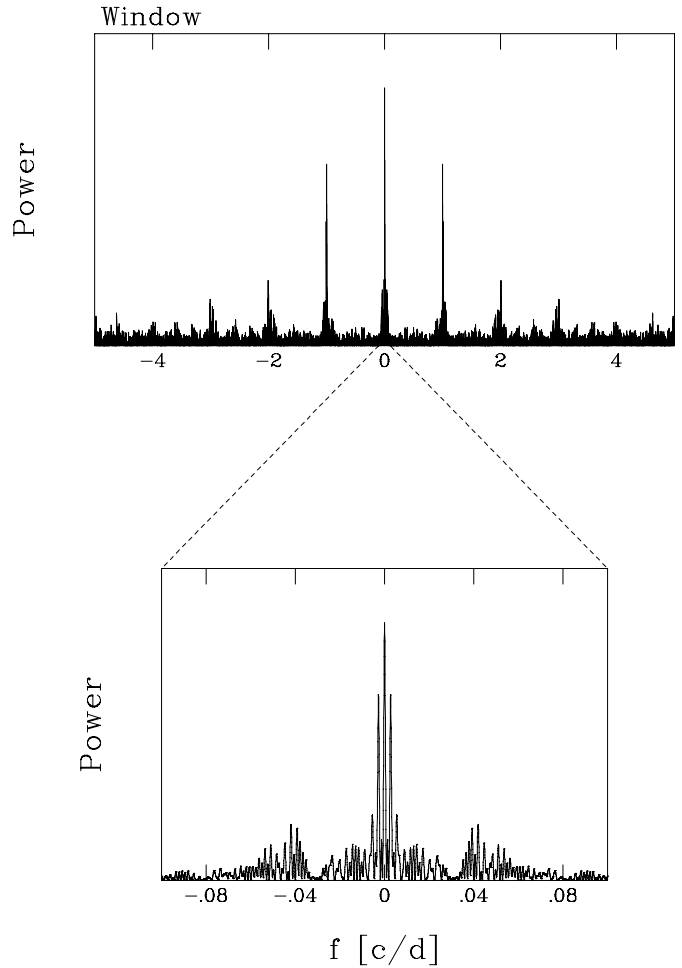}}
\caption{Spectral window for \object{BW6~V20} dataset.}
\label{one}
\end{figure}

\begin{figure}
\resizebox{\hsize}{!}{\includegraphics{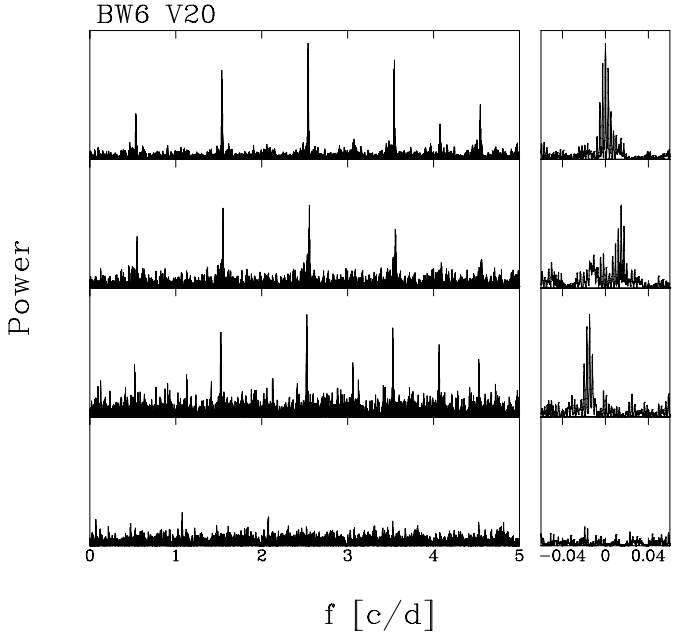}}
\caption{Frequency analysis of \object{BW6~V20}. Subsequent panels show power spectrum
         of the original data and the results of consecutive prewhitening 
         steps. Right column of the plot displays fine structure around the 
         main peak. The spectra are normalised separately in each panel.}
\label{two}
\end{figure}

\begin{figure}
\resizebox{\hsize}{!}{\includegraphics{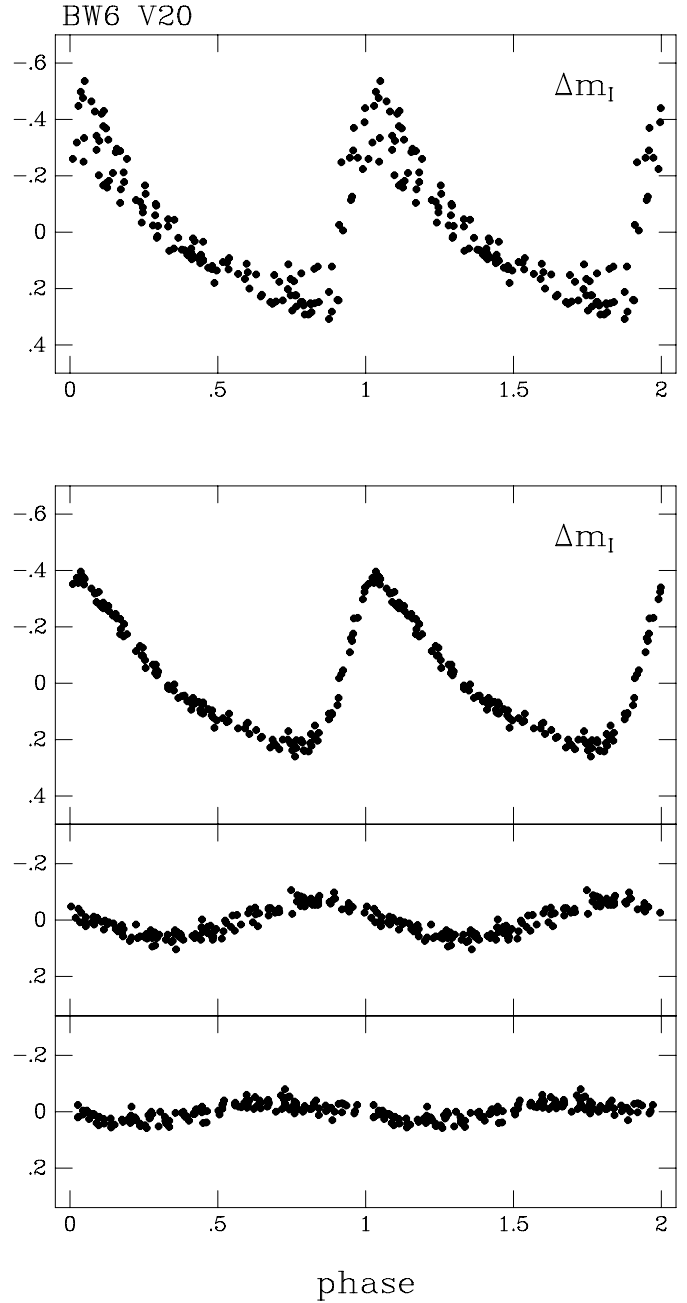}}
\caption{Phased lightcurve (I-magnitude) of \object{BW6~V20}. \emph{Top}: 
         original data phased with the primary period. \emph{Bottom}: 
         lightcurves of the three detected frequencies, 
         ${\rm f}_0 = 2.539947$\,c/d, ${\rm f}_1 = 2.554618$\,c/d and 
         ${\rm f}_2 = 2.525276$\,c/d, respectively.}
\label{three}
\end{figure}

As an example, we show application of the first procedure described
above to the case of \object{BW6~V20}. In Fig.\,\ref{one} we present the spectral
window of the data. This window is typical for OGLE-1 photometry.
Fig.\,\ref{two} shows consecutive steps of frequency analysis.  Top panel
displays Fourier power spectrum of the original data. The blow-up of the
neighborhood of the dominant peak is shown in the right column. The second
panel displays power spectrum of data prewhitened by the first frequency
and its harmonics. A new frequency has clearly emerged.  The next
prewhitening (panel three) reveals another frequency on the opposite side
of the dominant peak. Finally, after prewhitening by three frequencies no
more periodicities are detected (panel four). The analysis is stopped. In
Fig.\,\ref{three} we display the phased lightcurve of \object{BW6~V20}, first for the
original data (top panel), then separately for the three detected periods
(three bottom panels).

\section{Results}

Most of the RR~Lyrae stars in the OGLE-1 sample are strictly periodic. Those
stars are discussed in Paper~I. Departures from a monoperiodic
behaviour has been detected in 49 RR~Lyrae variables. They can be divided
in several distinctive types.

\subsection{RRd stars: variables with radial fundamental and first overtone 
            modes}

\begin{table}
\caption[]{RR-$\nu$1 variables of OGLE-1 sample}
\begin{tabular} {l r r  rr }
\hline
\hline
\noalign{\smallskip}
\multicolumn{1}{l}{OGLE No} & 
\multicolumn{1}{c}{${\rm P}_0$} &
\multicolumn{1}{c}{$\Delta{\rm f}$} & 
\multicolumn{1}{c}{${\rm A}_0$} &
\multicolumn{1}{c}{${\rm A}_1$}\\ 

\multicolumn{1}{c}{} & 
\multicolumn{1}{c}{[d]} &
\multicolumn{1}{c}{[c/d]} & 
\multicolumn{1}{c}{[mag]} &
\multicolumn{1}{c}{[mag]}\\ 
\noalign{\smallskip}
\hline
\noalign{\smallskip}
\multicolumn{5}{c}{RR1-$\nu$1} \\
\noalign{\smallskip}
\hline
\noalign{\smallskip}
  \object{BWC V47}       &  0.256926  &   --0.02439  &       0.067  &      0.050 \\
  \object{MM5A V20}$^{a}$&  0.391192  &     0.04546  &       0.126  &      0.009 \\
\noalign{\smallskip}
\hline
\noalign{\smallskip}
\multicolumn{5}{c}{RR0-$\nu$1}\\
\noalign{\smallskip}
\hline
\noalign{\smallskip}
  \object{MM5A V46}$^{a}$&  0.370187  &     0.02530  &       0.202  &      0.041 \\
  \object{BW3 V17}       &  0.403331  &     0.12005  &       0.258  &      0.013 \\
  \object{BW11 V10}      &  0.450658  &     0.01864  &       0.252  &      0.049 \\
  \object{BW3 V13}       &  0.457517  &     0.03255  &       0.230  &      0.042 \\
  \object{BWC V41}       &  0.462144  &     0.01767  &       0.221  &      0.041 \\
  \object{BW11 V44}      &  0.482493  &     0.03329  &       0.204  &      0.028 \\
  \object{BW5 V34}       &  0.490265  &   --0.00873  &       0.223  &      0.028 \\
  \object{BW2 V23}       &  0.492764  &     0.01198  &       0.244  &      0.044 \\
  \object{BW7 V15}$^{b}$ &  0.497071  &     0.00941  &       0.226  &      0.035 \\
  \object{BW7 V8}        &  0.507103  &   --0.03312  &       0.244  &      0.024 \\
  \object{BW7 V33}       &  0.511226  &   --0.00689  &       0.261  &      0.030 \\
  \object{MM5A V9}       &  0.516204  &     0.03580  &       0.251  &      0.016 \\
  \object{BW10 V21}      &  0.521996  &     0.01740  &       0.136  &      0.042 \\
  \object{MM5B V4}       &  0.524886  &     0.02237  &       0.184  &      0.024 \\
  \object{BW10 V20}      &  0.527123  &   --0.02450  &       0.244  &      0.023 \\
  \object{BW10 V44}      &  0.527263  &   --0.01206  &       0.191  &      0.036 \\
  \object{BW8 V15}       &  0.577234  &     0.02543  &       0.167  &      0.028 \\
  \object{BW5 V36}$^{c}$ &  0.594501  &     0.01445  &       0.139  &      0.015 \\
  \object{BW2 V24}       &  0.597426  &     0.01641  &       0.118  &      0.018 \\
  \object{BWC V61}       &  0.615946  &     0.02145  &       0.096  &      0.022 \\
  \object{BW10 V41}      &  0.626882  &     0.00855  &       0.215  &      0.049 \\
  \object{BW1 V34}       &  0.632719  &     0.01434  &       0.154  &      0.031 \\
  \object{BWC V51}       &  0.649494  &     0.01618  &       0.071  &      0.031 \\
  \object{BW6 V17}       &  0.651601  &     0.00545  &       0.146  &      0.022 \\
  \object{BW10 V40}      &  0.682741  &     0.01393  &       0.117  &      0.021 \\
\noalign{\smallskip}
\hline
\noalign{\smallskip}
\noalign{$^a$  Amplitude and/or phase of primary peak variable.} 
\noalign{$^b$  Second harmonic of primary frequency variable.}
\noalign{$^c$  Amplitude and/or phase of secondary peak variable.}
\end{tabular}
\label{df}
\end{table}

\begin{table*}
\centering
\caption[]{Amplitudes of Fourier components for selected RR-$\nu$1 variables 
           of OGLE-1 sample}
\begin{tabular} {l ccccc ccccc }
\hline
\hline
\noalign{\smallskip}
Frequency                     & \object{BWC V47}  
                                         & \object{MM5A V46}
                                                    & \object{BW11 V10}
                                                               & \object{BWC V41}
                                                                         & \object{BW3 V13}
                                                                                    & \object{BW5 V34}
                                                                                              & \object{MM5A V9} \\
\noalign{\smallskip}
\hline
\noalign{\smallskip}
\,\ $\Delta{\rm f}$           &   0.005  &    ---   &   0.014  &   ---   &    ---   &   ---   &   ---   \\
\noalign{\smallskip}
\,\ ${\rm f}_0$               &   0.067  &   0.202  &   0.252  &  0.221  &   0.230  &  0.223  &  0.251  \\
\,\ ${\rm f}_0+\Delta{\rm f}$ &   0.050  &   0.041  &   0.049  &  0.041  &   0.042  &  0.028  &  0.016  \\
\noalign{\smallskip}
$2{\rm f}_0$                  &   0.008  &   0.068  &   0.115  &  0.087  &   0.103  &  0.107  &  0.123  \\
$2{\rm f}_0+\Delta{\rm f}$    &    ---   &   0.028  &   0.040  &  0.032  &   0.027  &  0.025  &  0.010  \\
\noalign{\smallskip}
$3{\rm f}_0$                  &          &   0.021  &   0.072  &  0.044  &   0.062  &  0.072  &  0.080  \\
$3{\rm f}_0+\Delta{\rm f}$    &          &   0.011  &   0.033  &  0.022  &   0.022  &  0.024  &   ---   \\
\noalign{\smallskip}
$4{\rm f}_0$                  &          &          &   0.043  &  0.028  &   0.031  &  0.038  &  0.052  \\
$4{\rm f}_0+\Delta{\rm f}$    &          &          &   0.028  &   ---   &    ---   &   ---   &   ---   \\
\noalign{\smallskip}
$5{\rm f}_0$                  &          &          &   0.027  &         &          &  0.021  &  0.031  \\
$5{\rm f}_0+\Delta{\rm f}$    &          &          &   0.024  &         &          &   ---   &   ---   \\
\noalign{\smallskip}
$6{\rm f}_0$                  &          &          &   0.018  &         &          &  0.016  &  0.017  \\
$6{\rm f}_0+\Delta{\rm f}$    &          &          &   0.018  &         &          &   ---   &   ---   \\
\noalign{\smallskip}
$7{\rm f}_0$                  &          &          &          &         &          &         &  0.010  \\
$7{\rm f}_0+\Delta{\rm f}$    &          &          &          &         &          &         &   ---   \\
\noalign{\smallskip}
\hline
\hline
\noalign{\smallskip}
                              & \object{BW10 V21}
                                         & \object{BW10 V20}
                                                    & \object{BW10 V44}
                                                               & \object{BWC V61}
                                                                         & \object{BW10 V41}
                                                                                    & \object{BWC V51}
                                                                                              & \object{BW6 V17} \\
\noalign{\smallskip}
\hline
\noalign{\smallskip}
\,\ $\Delta{\rm f}$           &    ---   &    ---   &    ---   &   ---   &    ---   &   ---   &   ---   \\
\noalign{\smallskip}
\,\ ${\rm f}_0$               &   0.136  &   0.244  &   0.191  &  0.096  &   0.215  &  0.071  &  0.146  \\
\,\ ${\rm f}_0+\Delta{\rm f}$ &   0.042  &   0.023  &   0.036  &  0.022  &   0.049  &  0.031  &  0.022  \\
\noalign{\smallskip}
$2{\rm f}_0$                  &   0.066  &   0.118  &   0.083  &  0.034  &   0.100  &  0.019  &  0.068  \\
$2{\rm f}_0+\Delta{\rm f}$    &   0.029  &   0.022  &   0.024  &  0.014  &   0.032  &  0.009  &  0.010  \\
\noalign{\smallskip}
$3{\rm f}_0$                  &   0.033  &   0.062  &   0.052  &  0.015  &   0.052  &  0.006  &  0.041  \\
$3{\rm f}_0+\Delta{\rm f}$    &   0.023  &   0.024  &   0.008  &  0.016  &   0.029  &   ---   &  0.017  \\
\noalign{\smallskip}
$4{\rm f}_0$                  &   0.013  &   0.036  &   0.028  &         &   0.035  &         &  0.019  \\
$4{\rm f}_0+\Delta{\rm f}$    &    ---   &   0.016  &   0.021  &         &    ---   &         &  0.010  \\
\noalign{\smallskip}
$5{\rm f}_0$                  &          &   0.017  &          &         &          &         &  0.013  \\
$5{\rm f}_0+\Delta{\rm f}$    &          &   0.015  &          &         &          &         &   ---   \\
\noalign{\smallskip}
$6{\rm f}_0$                  &          &          &          &         &          &         &  0.007  \\
$6{\rm f}_0+\Delta{\rm f}$    &          &          &          &         &          &         &   ---   \\
\noalign{\smallskip}
\hline
\end{tabular}
\label{ct}
\end{table*}

This type of oscillations has been detected in one object of the sample:
\object{BW7~V30}. The star pulsates with ${\rm P}_{\! 0} = 0\fd 486889\pm 0\fd 000010$ 
and ${\rm P}_{\! 1} = 0\fd 36202176\pm 0\fd 00000083$. The resultant period 
ratio of ${\rm P}_{\! 1}/{\rm P}_{\! 0} = 0.743541\pm 0.000016$ is typical for the RRd
variables.  The first overtone strongly dominates over the fundamental mode,
with ${\rm A}_0/{\rm A}_1 = 0.142$.  Two harmonics of the first
overtone are detected, but no harmonics of the fundamental and no frequency
combinations can be identified. This is perhaps not surprising, considering
extremely low amplitude of the fundamental mode (${\rm A}_0 = 0.018$\,mag).

\subsection{RR-$\nu$1 stars: Blazhko variables with two closely spaced 
            frequencies}

In 38 RR~Lyrae variables secondary peaks close to the dominant pulsation
frequency are detected.  They are well-resolved within our dataset and are
not due to a secular period variability. Their beating with the primary
(radial) pulsation results in an apparent long-term amplitude and phase
modulation, a phenomenon referred to as Blazhko effect.

\begin{table}
\centering
\caption[]{RR-BL variables of OGLE-1 sample}
\begin{tabular} {l r r  rr }
\hline
\hline
\noalign{\smallskip}
\multicolumn{1}{l}{} & 
\multicolumn{1}{c}{} &
\multicolumn{1}{c}{$\Delta{\rm f}_{-}$} & 
\multicolumn{1}{c}{} &
\multicolumn{1}{c}{${\rm A}_{-}$}\\ 

\multicolumn{1}{l}{OGLE No} & 
\multicolumn{1}{c}{${\rm P}_0$} &
\multicolumn{1}{c}{$\Delta{\rm f}_{+}$} & 
\multicolumn{1}{c}{${\rm A}_0$} &
\multicolumn{1}{c}{${\rm A}_{+}$}\\ 

\multicolumn{1}{l}{} & 
\multicolumn{1}{c}{[d]} &
\multicolumn{1}{c}{[c/d]} & 
\multicolumn{1}{c}{[mag]} &
\multicolumn{1}{c}{[mag]}\\ 
\noalign{\smallskip}
\hline
\noalign{\smallskip}
\multicolumn{5}{c}{RR1-BL} \\
\noalign{\smallskip}
\hline
\noalign{\smallskip}
  \object{BW8 V34}  &     0.320365   &      0.01544   &        0.127  &       0.030 \\
                    &                &                &               &       0.016 \\
\noalign{\smallskip}
\hline
\noalign{\smallskip}
\multicolumn{5}{c}{RR0-BL} \\
\noalign{\smallskip}
\hline
\noalign{\smallskip}
  \object{BW6 V20}  &     0.393709   &      0.01467   &        0.244  &       0.031 \\
                    &                &                &               &       0.065 \\
\noalign{\smallskip}
  \object{BW9 V34}  &     0.449297   &      0.00896   &        0.254  &       0.042  \\
                    &                &                &               &       0.063  \\
\noalign{\smallskip}
  \object{BWC V15}  &     0.458717   &      0.00376   &        0.228  &       0.072 \\
                    &                &                &               &       0.056 \\
\noalign{\smallskip}
  \object{BW9 V24}  &     0.476339   &    --0.00185   &        0.206  &       0.021 \\
                    &                &      0.00239   &               &       0.103 \\
\noalign{\smallskip}
  \object{BW6 V7}   &     0.525015   &      0.03363   &        0.257  &       0.035 \\
                    &                &                &               &       0.042 \\
\noalign{\smallskip}
  \object{BW1 V18}  &     0.529560   &      0.01876   &        0.250  &       0.032 \\
                    &                &                &               &       0.034 \\
\noalign{\smallskip}
  \object{BWC V33}  &     0.550304   &      0.00865   &        0.175  &       0.020 \\
                    &                &                &               &       0.034 \\
\noalign{\smallskip}
  \object{BW6 V29}  &     0.562906   &      0.02283   &        0.164  &       0.019 \\
                    &                &                &               &       0.022 \\
\noalign{\smallskip}
  \object{BW10 V66} &     0.581234   &      0.01410   &        0.156  &       0.030 \\
                    &                &                &               &       0.053 \\
\noalign{\smallskip}
\hline
\end{tabular}
\label{bl}
\end{table}

\begin{table*}
\centering
\caption[]{Amplitudes of Fourier components for RR-BL variables of OGLE-1 
           sample}
\begin{tabular} {l ccccccccc}
\hline
\hline
\noalign{\smallskip}
Frequency                     & \object{BW8 V34}
                                        & \object{BW6 V20}
                                                  & \object{BW9 V34}
                                                            & \object{BWC V15}
                                                                      & \object{BW9 V24}
                                                                                & \object{BW6 V7}
                                                                                         & \object{BW1 V18}
                                                                                                   & \object{BWC V33}
                                                                                                             & \object{BW10 V66} \\
\noalign{\smallskip}
\hline
\noalign{\smallskip}
\,\ $\Delta{\rm f}$           &   ---   &   ---   &   ---   &   ---   &   ---   &   ---  &  0.020  &   ---   &    ---   \\
\noalign{\smallskip}
\,\ ${\rm f}_0-\Delta{\rm f}$ &  0.030  &  0.031  &  0.042  &  0.072  &  0.020  &  0.035 &  0.032  &  0.020  &   0.030  \\
\,\ ${\rm f}_0$               &  0.127  &  0.244  &  0.254  &  0.228  &  0.206  &  0.257 &  0.250  &  0.175  &   0.156  \\
\,\ ${\rm f}_0+\Delta{\rm f}$ &  0.016  &  0.065  &  0.063  &  0.056  &  0.103  &  0.042 &  0.034  &  0.034  &   0.053  \\
\noalign{\smallskip}
$2{\rm f}_0-\Delta{\rm f}$    &  0.008  &  0.028  &  0.039  &  0.059  &   ---   &  0.034 &  0.028  &  0.019  &   0.027  \\
$2{\rm f}_0$                  &  0.014  &  0.104  &  0.103  &  0.081  &  0.068  &  0.124 &  0.127  &  0.087  &   0.062  \\
$2{\rm f}_0+\Delta{\rm f}$    &   ---   &  0.042  &  0.051  &  0.044  &  0.066  &  0.038 &  0.020  &  0.033  &   0.035  \\
\noalign{\smallskip}
$3{\rm f}_0-\Delta{\rm f}$    &         &  0.021  &  0.030  &  0.035  &   ---   &  0.029 &  0.019  &   ---   &   0.016  \\
$3{\rm f}_0$                  &         &  0.040  &  0.052  &  0.036  &  0.018  &  0.065 &  0.080  &  0.051  &   0.036  \\
$3{\rm f}_0+\Delta{\rm f}$    &         &  0.030  &  0.050  &  0.024  &  0.051  &  0.038 &   ---   &  0.030  &   0.028  \\
\noalign{\smallskip}
$4{\rm f}_0-\Delta{\rm f}$    &         &  0.012  &  0.022  &  0.016  &   ---   &  0.024 &  0.018  &   ---   &    ---   \\
$4{\rm f}_0$                  &         &  0.022  &  0.020  &  0.022  &   ---   &  0.038 &  0.049  &  0.033  &   0.017  \\
$4{\rm f}_0+\Delta{\rm f}$    &         &  0.029  &  0.027  &  0.014  &  0.026  &  0.030 &   ---   &  0.023  &   0.019  \\
\noalign{\smallskip}
$5{\rm f}_0-\Delta{\rm f}$    &         &   ---   &         &  0.011  &         &  0.015 &   ---   &         &    ---   \\
$5{\rm f}_0$                  &         &  0.013  &         &  0.017  &         &  0.018 &  0.020  &         &    ---   \\
$5{\rm f}_0+\Delta{\rm f}$    &         &  0.015  &         &  0.013  &         &  0.022 &   ---   &         &   0.014  \\
\noalign{\smallskip}
\hline
\end{tabular}
\label{abl}
\end{table*}

In most cases (27 stars) only one secondary peak is present, forming a
close \emph{doublet} with the primary peak. Such frequency pattern occurs in
2~RRc and in 25~RRab stars.  Following Alcock et~al. (\cite{alc00}), we denote
these variables as RR1-$\nu$1 and RR0-$\nu$1, respectively. Their main
properties (period ${\rm P}_{\! 0}$ and amplitude ${\rm A}_0$ of the
primary peak, frequency separation $\Delta {\rm f} = {\rm f}_1 - {\rm f}_0$
and amplitude of the secondary peak ${\rm A}_1$) are listed in
Table~\ref{df}. In most stars the frequency separation is between 
0.008\,c/d and 0.036\,c/d, corresponding to a beat (Blazhko) period of 
$28\! -\! 125$\,day. The amplitude of the secondary peak is always very small, 
never exceeding 0.05\,mag.  Contrary to the case of the LMC stars (Alcock
et~al. \cite{alc00}; Welch et~al. \cite{wel02}), combination frequencies 
$k{\rm f}_0 + {\rm f}_1 = (k+1){\rm f}_0 + \Delta{\rm f}$ can be detected in 
about half of the RR-$\nu$1 stars of the OGLE-1 sample. The amplitudes of
all components identified in these variables are given in Table~\ref{ct}.
The amplitudes and frequencies listed in Tables~\ref{df} and \ref{ct} are
determined through the least--squares fit of Eq.\,(2) to the data.

\subsection{RR-BL stars: Blazhko variables with three symmetrically spaced 
            close frequencies}

\begin{table}
\centering
\caption[]{RR0-$\nu$2 variable of OGLE-1 sample}
\begin{tabular} {l r r  rr }
\hline
\hline
\noalign{\smallskip}
\multicolumn{1}{l}{OGLE No} & 
\multicolumn{1}{c}{${\rm P}_0$} &
\multicolumn{1}{c}{$\Delta{\rm f}$} & 
\multicolumn{1}{c}{${\rm A}_0$} &
\multicolumn{1}{c}{${\rm A}_1$}\\ 

\multicolumn{1}{c}{} & 
\multicolumn{1}{c}{[d]} &
\multicolumn{1}{c}{[c/d]} & 
\multicolumn{1}{c}{[mag]} &
\multicolumn{1}{c}{[mag]}\\ 
\noalign{\smallskip}
\hline
\noalign{\smallskip}
  \object{BWC V106} &      0.464967  &     0.00708     &    0.188    &    0.020 \\
                    &                &     0.02675     &             &    0.020 \\
\noalign{\smallskip}
\hline
\end{tabular}
\label{rrd}
\end{table}

In several RR~Lyrae variables \emph{two} secondary frequencies, located on
opposite sides of the primary peak, are present. Together with the primary
frequency they form a close \emph{equally spaced triplet}. Such frequency
pattern is found in 1~RRc and in 9~RRab stars. Following Alcock et~al.
(\cite{alc00}), we denote these variables as RR1-BL and RR0-BL, respectively. Their
properties are listed in Table~\ref{bl}. The amplitudes of secondary peaks are
somewhat larger than in the RR-$\nu$1 stars, reaching up to 0.1\,mag.  The
frequency separations $\Delta{\rm f}$ are similar to those observed in
RR-$\nu$1 stars.  With the exception of one star, the triplets are 
\emph{exactly equidistant}.  This has been verified by fitting the data with
Eq.\,(2), with all 3 frequencies of a triplet treated as independent
parameters.  The resulting separations $-\Delta{\rm f}_{-} = -({\rm f}_{-}
- {\rm f}_0)$ and $\Delta{\rm f}_{+} = {\rm f}_{+} - {\rm f}_0$ differ by
less than 0.00005\,c/d, which within our dataset is indistinguishable from
zero. Thus, in the final solution $\Delta{\rm f}_{-} = -\Delta{\rm f}_{+}$ is
assumed. The one deviating case is \object{BW9~V24}, where $\Delta{\rm f}_{-} +
\Delta{\rm f}_{+} = 0.00054$\,c/d. We note, that in this particular star the 
triplet is barely resolved. Since the ${\rm A}_{-}$ component in \object{BW9~V24} is
very weak, the insufficient resolution can have a significant effect on its
frequency determination.  The deviation from equidistant spacing seen in
\object{BW9~V24} needs to be confirmed with a longer dataset.

\begin{figure*}
\resizebox{\hsize}{!}{\includegraphics{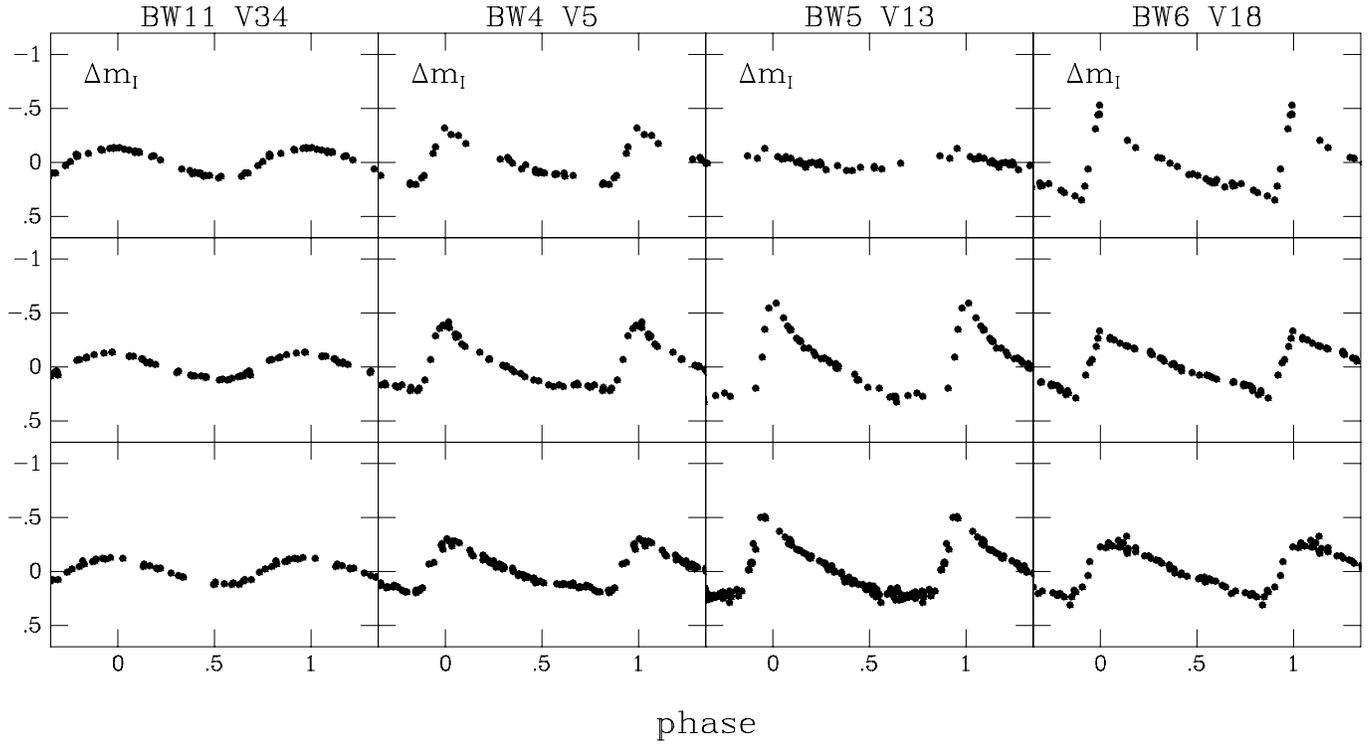}}
\caption{Seasonal lightcurves (I-magnitude) of suspected Blazhko 
         variables. Panels from top to bottom correspond to observing seasons
         of 1992, 1993 and 1994, respectively.}
\label{four}
\end{figure*}

In all but one RR-BL stars we detect combination frequencies
$k{\rm f}_0\pm\Delta{\rm f}$. The amplitudes of all identified components
are given in Table~\ref{abl}. The combination peaks, together with harmonics of
${\rm f}_0$, form a sequence of equidistant triplets.  This structure can be
followed up to the third or fourth harmonic. We note in passing, that the
observed frequency pattern is the same as found in several well-studied
field RR~Lyrae Blazhko variables (Borkowski \cite{bor80}; 
Smith et~al. \cite{smi94}, \cite{smi99};
Kov\'acs \cite{kov95}; Nagy \cite{nag98}; Szeidl \& Koll\'ath \cite{sze00}).

\begin{table}
\centering
\caption[]{Variable period RRc stars of OGLE-1 sample}
\begin{tabular} {l r r  rr }
\hline
\hline
\noalign{\smallskip}
\multicolumn{1}{l}{OGLE No} & 
\multicolumn{1}{c}{${\rm P}_0$} &
\multicolumn{1}{c}{${\rm A}_0$} & 
\multicolumn{1}{c}{$\dot{P}$} &
\multicolumn{1}{c}{P/$\dot{P}$}\\ 

\multicolumn{1}{c}{} & 
\multicolumn{1}{c}{[d]} &
\multicolumn{1}{c}{[mag]} & 
\multicolumn{1}{c}{[s/s]} &
\multicolumn{1}{c}{[yr]}\\
\noalign{\smallskip}
\hline 
\noalign{\smallskip}
  \object{BWC V37}   & 0.380122 & 0.140 &   1.7$\times 10^{-7}$ &   6.1$\times 10^3$ \\
  \object{BW2 V10}   & 0.507782 & 0.118 &   3.1$\times 10^{-7}$ &   4.5$\times 10^3$ \\
  \object{BW7 V51}   & 0.272148 & 0.154 &   2.9$\times 10^{-8}$ &   2.5$\times 10^4$ \\
  \object{BW9 V38}   & 0.305731 & 0.117 & --4.0$\times 10^{-8}$ & --2.1$\times 10^4$ \\
  \object{BW11 V55}  & 0.286179 & 0.123 & --5.1$\times 10^{-8}$ & --1.5$\times 10^4$ \\
  \object{MM5A V24}  & 0.379396 & 0.093 & --5.4$\times 10^{-8}$ & --1.9$\times 10^4$ \\
\noalign{\smallskip}
\hline
\end{tabular}
\label{pdot}
\end{table}

\begin{table}
\centering
\caption[]{Suspected Blazhko variables of OGLE-1 sample}
\begin{tabular} {l r r  rr }
\hline
\hline
\noalign{\smallskip}
\multicolumn{1}{l}{OGLE No} & 
\multicolumn{1}{c}{${\rm P}_0$} &
\multicolumn{1}{c}{${\rm A}_1$} & 
\multicolumn{1}{c}{$\phi_1$}\\ 

\multicolumn{1}{c}{} & 
\multicolumn{1}{c}{[d]} &
\multicolumn{1}{c}{[mag]} & 
\multicolumn{1}{c}{[rad]}\\ 
\noalign{\smallskip}
\hline
\noalign{\smallskip}
\multicolumn{4}{c}{RRc}\\
\noalign{\smallskip}
\hline
\noalign{\smallskip}
  \object{BW11 V34}   &   0.341674 &   $0.139\pm 0.002$ &   $0.00\pm 0.02$ \\
                      &            &   $0.122\pm 0.002$ &   $0.14\pm 0.02$ \\
                      &            &   $0.119\pm 0.003$ &   $0.03\pm 0.02$ \\
\noalign{\smallskip}
\hline
\noalign{\smallskip}
\multicolumn{4}{c}{RRab}\\
\noalign{\smallskip}
\hline
\noalign{\smallskip}
  \object{BW4 V5}     &   0.474673 &   $0.167\pm 0.013$ &   $0.00\pm 0.05$ \\
                      &            &   $0.223\pm 0.004$ &   $0.13\pm 0.02$ \\
                      &            &   $0.182\pm 0.003$ & --$0.04\pm 0.02$ \\
\noalign{\medskip}
  \object{BW5 V13}    &   0.494952 &   $0.065\pm 0.006$ &   $0.00\pm 0.10$ \\
                      &            &   $0.326\pm 0.006$ & --$1.42\pm 0.04$ \\
                      &            &   $0.286\pm 0.005$ & --$1.01\pm 0.02$ \\
\noalign{\medskip}
  \object{BW6 V18}    &   0.541402 &   $0.290\pm 0.014$ &   $0.00\pm 0.03$ \\
                      &            &   $0.199\pm 0.004$ & --$0.22\pm 0.02$ \\
                      &            &   $0.205\pm 0.005$ & --$0.24\pm 0.03$ \\
\noalign{\smallskip}
\hline
\end{tabular}
\label{blaz}
\end{table}

\subsection{RR-$\nu$2 stars: Blazhko variables with three nonequidistant 
            close frequencies}

Yet another pulsation patter is found in \object{BWC~V106}. In this star, two
secondary frequencies are detected, but both located on \emph{the same side}
of the primary peak. Together with the primary frequency they form a close
triplet, which is \emph{neither equally spaced, nor centered on the primary}.
Following Alcock et~al. (\cite{alc00}) we clasify \object{BWC~V106} as RR0-$\nu$2-type
variable. This type of pulsation pattern is extremely rare, only 3 similar
stars have been identified among 1327 first overtone RR~Lyrae pulsators in
the LMC (Alcock et~al. \cite{alc00}). The properties of \object{BWC~V106} are listed in
Table~\ref{rrd}.  No combination peaks are detected in this star.

\subsection{Miscellaneous RR~Lyrae variables}

\begin{table*}
\centering
\caption[]{Variable Types in OGLE-1 Galactic Bulge Sample: 
           First overtone RR~Lyrae Stars (RRc Stars)}
\begin{tabular} {l l c  rr }
\hline
\hline
\noalign{\smallskip}
\multicolumn{1}{l}{Type} & 
\multicolumn{1}{c}{Description} &
\multicolumn{1}{c}{Number} & 
\multicolumn{1}{c}{\%} &
\multicolumn{1}{c}{LMC \%}\\ 
\noalign{\smallskip}
\hline
\noalign{\smallskip}
  RRc \dotfill            & single-mode          &  54 & $83.1\pm 11.3$     & $69.0\pm 1.3$ \\
  RRd\,\, \dotfill        & F/1H double mode     & ~~1 &  $1.5\pm \ \, 1.5$ & $13.6\pm 0.9$ \\
  RR1-$\nu$1\,\, \dotfill & two close components & ~~2 &  $3.1\pm \ \, 2.2$ &  $1.8\pm 0.4$ \\
  RR1-BL\, \dots\dots     & symmetric triplet    & ~~1 &  $1.5\pm \ \, 1.5$ &  $2.1\pm 0.4$ \\
  RRc-PC\, \dots\dots     & period change        & ~~6 &  $9.2\pm \ \, 3.8$ & $10.6\pm 0.9$ \\
  RRc-NC\, \dots\dots     & nonclassified        & ~~1 &  $1.5\pm \ \, 1.5$ &  $0.5\pm 0.2$ \\
\noalign{\smallskip}
\hline
\end{tabular}
\label{irc}
\end{table*}

\begin{table*}
\centering
\caption[]{Variable Types in OGLE-1 Galactic Bulge Sample:
           Fundamental Mode RR~Lyrae Stars (RRab Stars)}
\begin{tabular} {l l c  rr }
\hline
\hline
\noalign{\smallskip}
\multicolumn{1}{l}{Type} & 
\multicolumn{1}{c}{Description} &
\multicolumn{1}{c}{Number} & 
\multicolumn{1}{c}{\%} &
\multicolumn{1}{c}{LMC \%}\\ 
\noalign{\smallskip}
\hline
\noalign{\smallskip}
  RRab \dotfill         & single-mode            &    112 & $74.7\pm 7.1$  & ?~~~~~       \\
  RR0-$\nu$1 \dotfill   & two close components   &   ~~25 & $16.7\pm 3.3$  & $6.5\pm 0.7$ \\
  RR0-$\nu$2 \dotfill   & three close components &  ~~~~1 &  $0.7\pm 0.7$  & ?~~~~~       \\
  RR0-BL\, \dotfill     & symmetric triplet      &  ~~~~9 &  $6.0\pm 2.0$  & $3.7\pm 0.5$ \\
  RRab-PC\, \dots\dots  & period change          & ~~~~-- &     ------~~~  & ?~~~~~       \\
  RRab-NC\, \dots\dots  & nonclassified          &  ~~~~3 &  $2.0\pm 1.2$  & ?~~~~~       \\
\noalign{\smallskip}
\hline
\end{tabular}
\label{irab}
\end{table*}

In 10 RR~Lyrae stars we find after prewhitening a significant residual
power at frequency almost identical to that of the (just removed) primary
component. Such behaviour is a signature of a slow phase and/or amplitude
variability, not resolved within the lenght of available data 
($\sim 900$~days for OGLE-1).

For closer examination of the long-term variations in these objects we have
performed Fourier fits of their lightcurves \emph{separately for each
observing season}. In 6 stars, all pulsating in the first overtone, we find
the pulsation phase to vary from year to year, but the amplitude of the
pulsation remains constant. These stars are listed in Table~\ref{pdot}. 
Clearly, they undergo a secular period change. Assuming, that the period 
varies linearly in time, we can estimate for each object the \emph{slowest} 
$dP/dt$ consistent with the data. We show them in the last two columns of 
Table~\ref{pdot}. We stress, that the numbers derived here are based on three
observing seasons only and therefore should be treated as preliminary.
Nevertheless, we can conclude that in all 6 cases pulsation period changes
on a timescale of no more than $\sim 10^4$\,yr, which is much shorter
than the timescale predicted by the theory of stellar evolution.

In the remaining 4 stars (1 RRc and 3 RRab-type) \emph{both} amplitude
and phase vary from year to year. We list these objects in Table~\ref{blaz}, 
where in the last two columns we give the values of the Fourier amplitude
${\rm A}_1$ and phase $\phi_1$ (see Eq.\,(1)) for each observing season.
Seasonal lightcurves of the stars are displayed in Fig.\,\ref{four}. The
amplitude variability is hardly noticable in the RRc star \object{BW11~V34}, but in
the RRab stars it is rather obvious.  In the extreme case of \object{BW5~V13} the
amplitude increases by factor of five within one year. The change of the
pulsation amplitude is accompanied by change of phase and change of shape
of the lightcurve . This last effect is most noticable in \object{BW4~V5} and
\object{BW6~V18}.  All 4 stars are most likely Blazhko variables with modulation
period (beat period between the modes) longer than the span of available
data. However, with the data in hand we are not able to make any firm
statement about the nature of the observed long term behaviour. In the
following discussion we will treat all 4 variables as "nonclassified".

\section{Discussion}

\subsection{Incidence rates}

Tables~\ref{irc} and \ref{irab} present the inventory of different types of
RR~Lyrae variables in the OGLE-1 Galactic Bulge sample, separately for the
overtone and for the fundamental mode pulsators. The canonical double mode
variables (RRd stars) are somewhat arbitrarily included into the table for
overtone variables. This is justified by strong dominance of the overtone
mode in these stars -- as a result they are usually disguised, until full
frequency analysis, as RRc stars. For each type of variability we list the
number of stars found (third column ) and the estimated incidence rate in
the population (fourth column).  The standard deviations of the incidence
rates are calculated assuming Poisson distribution. For the purpose of
comparison, we also show the corresponding rates for the LMC sample (Alcock
et~al. \cite{alc00}; Welch et~al. \cite{wel02}), these are listed in the last column of the
tables. The LMC rates have not been published for all the subclasses;
the missing entries in the tables are substituted with the question marks.

Several important conclusions can be drawn from data of Tables~\ref{irc} and 
\ref{irab}:

\begin{enumerate}
\item Blazhko variables (RR-$\nu$1, RR-$\nu$2, RR-BL) occur more 
                frequently among fundamental-mode pulsators (RRab) than 
                among overtone pulsators (RRc). The property is common for
                both Galactic Bulge and LMC populations, being somewhat 
                stronger in the Bulge. 

\item For the RRc stars, the fraction of Blazhko variables is 
                (within statistical error) the same in the Galactic Bulge and
                in the LMC. In case of the RRab stars, however, the fraction 
		is \emph{twice higher in the Bulge} ($23.3\pm 3.9$\% vs.
                $10.2\pm 0.8$\%). 

                This difference \emph{does not} result from lower noise 
                level of the Galactic Bulge data. The weakest secondary peaks 
                detected in the LMC stars have amplitudes of 
		$\sim 0.020$\,mag. Adopting the same threshold for our 
		sample, we are still left with 31 RRab Blazhko stars, which 
		is $20.7\pm 3.7\%$ of the RRab population. We conclude, that 
		the higher incidence rate of RRab Blazhko stars in the 
		Galactic Bulge, as compared to the LMC, is real. It is 
		tempting to speculate that this effect is caused by 
		difference in metallicity of the two populations. 

\item The RRd variables are \emph{significantly less common} in the 
                Galactic Bulge than in the LMC. Assuming the same incidence 
                rate as in the LMC, we would expect to find at least 7 
                variables of this type among 65 RRc stars of the Bulge 
                (3$\sigma$ lower limit). Instead, we find only one. Again, 
                different metallicity of the two populations is the likely 
                culprit.

\item Secular period change is detected in about 10\% of all 
                Galactic Bulge RRc stars. This is the same fraction as for the
                RRc stars in the LMC. Interestingly enough, no RRab stars in 
                the Bulge show this type of behaviour. We remind here, that 
                the observed period variability is too fast to be of 
                evolutionary nature. Its origin is currently unknown.
\end{enumerate}

\subsection{Pulsation properties of RRab Blazhko stars}

\begin{figure}
\resizebox{\hsize}{!}{\includegraphics{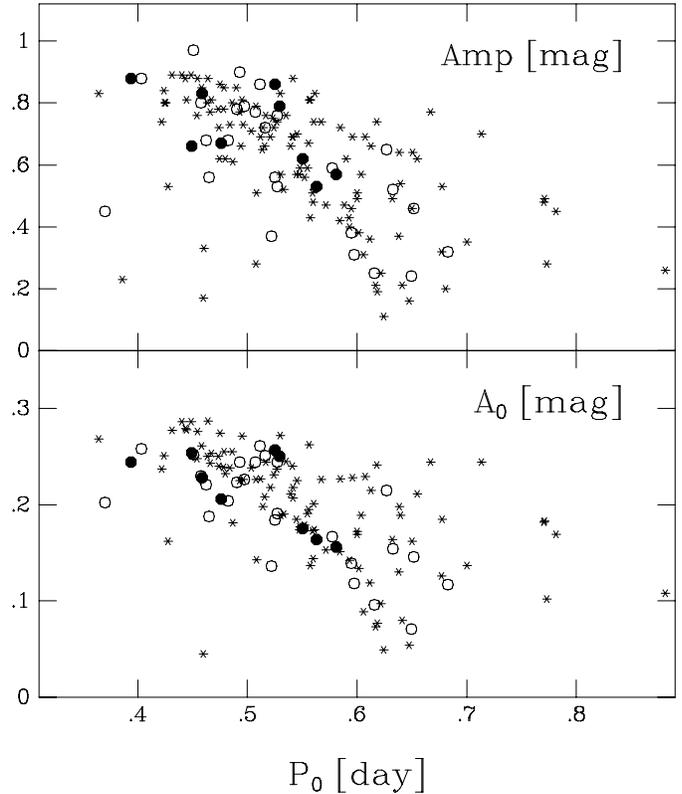}}
\caption{Period-Amplitude diagram for OGLE-1 RRab stars. \emph{Top}: 
         peak-to-peak amplitudes from OGLE-1 catalog, \emph{bottom}: Fourier 
         amplitudes of the radial mode. Monoperiodic variables marked by 
         asterisks, Blazhko variables by open circles (RR0-$\nu$1 and 
         RR0-$\nu$2) and filled circles (RR0-BL).}
\label{five}
\end{figure}

\begin{figure}
\resizebox{\hsize}{!}{\includegraphics{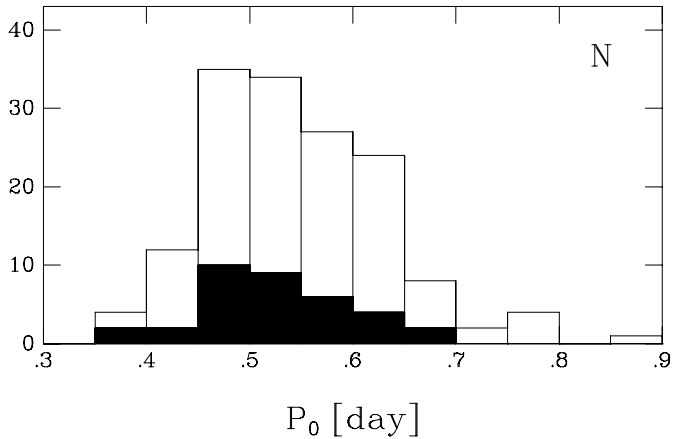}}
\caption{Period distribution of OGLE-1 RRab stars. Black area corresponds 
         to Blazhko variables.}
\label{six}
\end{figure}

\begin{figure}
\resizebox{\hsize}{!}{\includegraphics{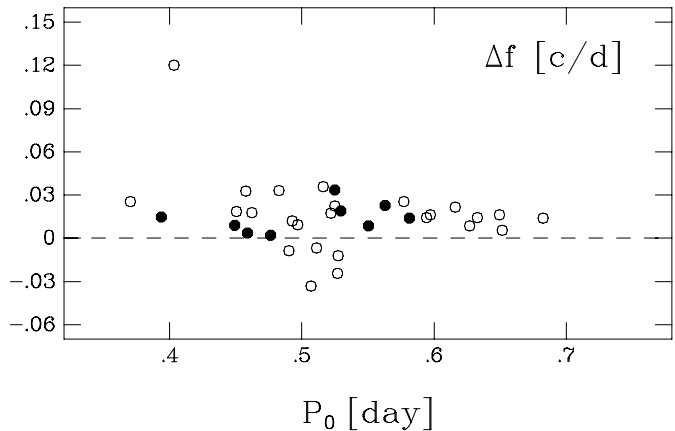}}
\caption{Frequency difference, $\Delta{\rm f}$, for OGLE-1 RRab Blazhko 
         variables. Same symbols as in Fig.\,\ref{five}.}
\label{seven}
\end{figure}

With the sample of 35 fundamental mode Blazhko stars, we are in position to
discuss the group properties of this type of variables. In Fig.\,\ref{five} we
display the Period--Amplitude diagram for the Galactic Bulge RRab stars.
In the upper panel we plot peak-to-peak amplitude, as given by OGLE-1
catalog. In the lower panel we plot the Fourier amplitude of the radial mode,
as determined by the Fourier fit Eq.\,(1) or Eq.\,(2).
Monoperiodic variables are marked by asterisks, the Blazhko variables by
open circles (RR0-$\nu$1 and RR0-$\nu$2) and by filled circles (RR0-BL).
It is immediately apparent that the presence of secondary frequencies
neither depends on nor affects the amplitude of the primary (radial)
pulsation. It does not depend on the pulsation period, either.  The stars with
close frequency doublets (RR0-$\nu$1 variables) are detected with roughly
the same probability at all periods represented in the sample.  The
occurrence of RR0-BL variables, however, seems to be limited to ${\rm P} <
0\fd 6$. Nevertheless, the overall fraction of Blazhko variables is
essentially period independent. This conclusion is also supported by the
histogram shown in Fig.\,\ref{six}. The fraction of Blazhko variables in each period
bin is within 1$\sigma$ consistent with the average value of 23.3\%.

In Fig.\,\ref{seven} we show the frequency separation $\Delta{\rm f} = {\rm f}_1 -
{\rm f}_0$ for RRab Blazhko variables of our sample. $\Delta{\rm f}$ displays
no trend with the radial mode period ${\rm P}_{\! 0}$. For 80\% of
RR0-$\nu$1 stars frequency separation is positive, corresponding to
secondary frequency being \emph{higher} than the primary one. Identical
distribution of $\Delta{\rm f}$ has also been found for RRab stars in the LMC
(Welch et~al. \cite{wel02}).  The negative values of $\Delta{\rm f}$ occur only in
the narrow range of periods between 0\fd 49 and 0\fd 53. Interestingly,
this particular period range is avoided by RR0-BL stars. The values of
$\Delta{\rm f}$ for triplets (RR0-BL stars) are on average slightly smaller
than for the doublets (RR0-$\nu$1 stars). In 8 out of 9 RR0-BL stars the
higher amplitude secondary peak has also higher frequency. Again, very
similar statistics has been found for the RRab stars in the LMC.

While $\Delta{\rm f}$ distribution for the RRab stars is very similar in both
galaxies, it is distinctively different than for the RRc stars of the LMC
(Alcock et~al. \cite{alc00}). In the latter case, the frequency separation is
negative in more than half of the doublets (RR1-$\nu$1 stars) and the
higher amplitude secondary peak in the triplets (RR1-BL stars) can occur on
either side of the primary peak, with about equal probability.  The values
of $|\Delta{\rm f}|$ in the LMC RRc stars are also noticably \emph{larger} 
than in the RRab stars.  Unfortunately, the number of RRc Blazhko variables
in the Galactic Bulge is too small for a meaningfull discussion.
Nevertheless, even for this very limited sample we find the same result:
average $\Delta{\rm f}$ is larger than for the RRab stars.  Apparently,
whatever mechanism is responsible for excitation of the secondary
frequencies, it works diferently in the overtone and in the fundamental
mode RR~Lyrae pulsators.

\subsection{Blazhko variables and metallicity}

\begin{figure}
\resizebox{\hsize}{!}{\includegraphics{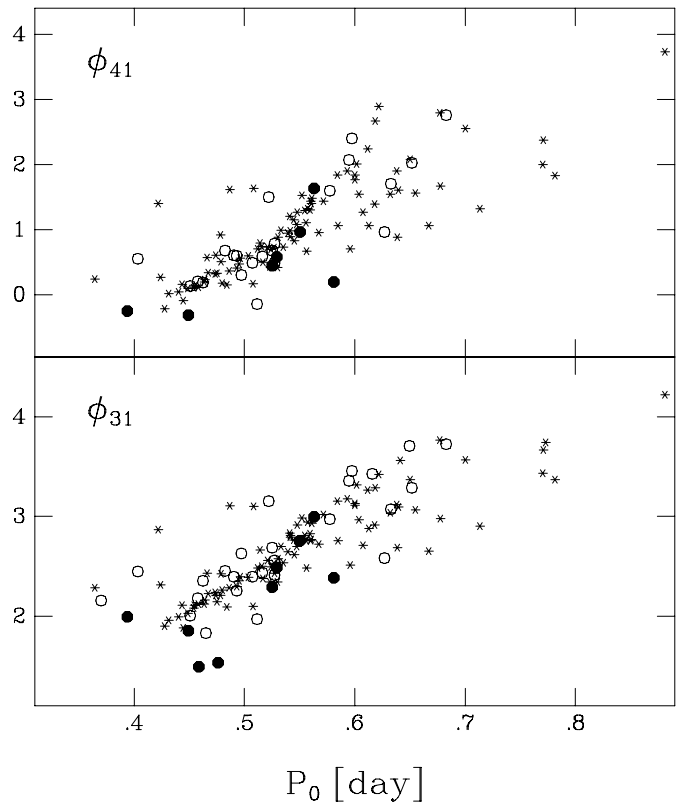}}
\caption{Fourier phases $\phi_{31}$ and $\phi_{41}$ for OGLE-1 RRab stars. 
         Same symbols as in Fig.\,\ref{five}.}
\label{eight}
\end{figure}

In Fig.\,\ref{eight} we display the Fourier phases $\phi_{31}$ and $\phi_{41}$ for all
the Galactic Bulge RRab stars. This plot for the monoperiodic pulsators
has already been presented in Paper~I (their Fig.\,12), here we have added
the multiperiodic (Blazhko) stars.  In case of multiperiodic stars, the
Fourier phases are derived for the primary (radial) component of the
pulsation.  Specifically, they are constructed with the phases of the 
$k{\rm f}_0$ components of the Fourier fit Eq.\,(2). We stress that this is
a different procedure than used by Jurcsik et al. (\cite{jur02}), who
have evaluated the Fourier parameters of the snapshot lightcurves at various
phases of the beat (Blazhko) period. As a result, their parameters are time
dependent.  The phases derived in the current paper do not vary in time.
They describe the shape of the primary (radial) pulsation \emph{after
secondary frequencies have been filtered out}.

In the case of  monoperiodic stars the Fourier phases $\phi_{31}$ and 
$\phi_{41}$ follow a single progression up to ${\rm P}\sim 0\fd 55$. At 
longer periods the progression splits into three separate tails. As has
been shown in Paper~I, these tails are formed by stars of different
metallicity, with the upper tail corresponding to the highest and the lower
tail to the lowest value of [Fe/H].

Fig.\,\ref{eight} shows that the Blazhko variables of our sample choose preferentially 
the \emph{upper tail} of the progression.  This is best visible in the plot
of $\phi_{31}$.  Out of 11 Blazhko stars with ${\rm P} > 0\fd 55$, 7 fall
on the upper tail, while only 2 belong to the central and to the lower
tails, each.  In the $\phi_{41}$ plot, these variables are split between
the tails \emph{in the same way}.  The distribution of Fourier phases
implies that the incidence rate of Blazhko variables increases with
metallicity.  We note, that the same dependence has already been suggested
by comparison between the Galactic Bulge and the LMC (see Sect.\,4.1).

\section{Conclusions}

We have conducted a systematic search for multiperiodic pulsators in the
OGLE-1 sample of Galactic Bulge RR~Lyrae stars. Multiperiodicity has been
established for 39 variables. In addition, we have identified 6 RRc stars
with changing pulsation period and 4 stars exhibiting long-term change of
pulsation amplitude.

Among 39 multiperiodic variables only one is a ``canonical" double-mode
pulsator (RRd stars), with two radial modes -- fundamental and first
overtone -- excited.  In the remaining 38 variables a different type of
multiperiodicity is found -- additional peaks are present very close to the
primary pulsation frequency. Their beating with the primary (radial)
pulsation results in an apparent long term amplitude and phase modulation.
This phenomenon is referred to as Blazhko effect. Majority of the Blazhko
variables come in one of two flavours: we detect either a single secondary
peak, forming a \emph{doublet} with the primary frequency (RR-$\nu$1 stars),
or a pair of secondary peaks, which together with the primary frequency
form an \emph{equidistant triplet} centered on the primary peak (RR-BL
stars).  While the frequency triplet can result from periodic amplitude
and/or phase modulation of a purely radial pulsation, such a process cannot
produce a doublet. The observed period ratios of ${\rm P}_{\! 1}/{\rm
P}_{\! 0} = 0.95-1.02$ are not compatible with the excitation of two radial
modes. This implies that a secondary component of the doublet must
correspond to a \emph{nonradial mode of oscillation}.

The Blazhko variables occur among both RRc (overtone) and RRab
(fundamental-mode) stars, but the incidence rate is not the same. In case
of the Galactic Bulge it is ~5\% for the RRc stars and ~23\% for the RRab
stars. 

The incidence rates of multiperiodic RR~Lyrae variables differ between
the Galactic Bulge and the Large Magellanic Cloud. Specifically, the
double-mode (RRd) stars occur much more frequently in the LMC, while the
opposite is true for the RRab Blazhko stars.  We believe that the effect is
caused by different metallicity of the two populations. Such interpretation
is supported by the Fourier analysis of the Bulge RRab stars. The
hypothesis of metallicity influence can be further tested when statistics of
multiperiodic variables in the SMC is established.

The occurence of Blazhko variability among RRab stars correlates neither
with amplitude nor with period of the primary pulsation. In vast majority
of doublets (RR0-$\nu$1 stars) the secondary frequency is higher than the
primary one. The same is also true for the higher of the two secondary peaks
in the triplets (RR0-BL stars). This is a different distribution than
observed in case of the RRc stars (Alcock et~al. \cite{alc00}). The values of the
frequency separation $\Delta{\rm f}$ are also different, being systematically
smaller than in the RRc stars. Clearly, the detailed picture of the Blazhko
variability \emph{depends on the primary mode of pulsations}.

The observational properties discussed in the current paper put constraints
on any proposed explanation of the Blazhko effect.  Two models are most
popular nowadays: the oblique magnetic pulsator model 
(Shibahashi \cite{shi95},
\cite{shi00}) and the 1:1 resonance model (Nowakowski \& Dziembowski \cite{now01}). The
results presented here seem to rule out the first of the two. The oblique
pulsator model predicts splitting pulsation frequency of a mode into
an equally spaced \emph{quintuplet}. We never detect such a structure,
despite specifically looking for it. We see only triplets or doublets. The
model cannot explain a striking difference of the Blazhko incidence rate
between RRc and RRab stars. If the modulation is indeed due to presence of
the magnetic field, then it should occur equally likely independently of
the pulsation mode. Finally, as the pulsation amplitude is supposed to vary
with the rotation period of the stars, it is hard to understand why Blazhko
periods of RRc stars are systematically shorter than that of the RRab
stars.

The resonant model avoids all of the above difficulties. It generates
equally split triplets, not quintuplets. It can explain naturally the
asymmetry of amplitudes in the triplet. The model predicts higher incidence
rate of Blazhko variables among RRab than among RRc stars (Dziembowski \&
Cassisi \cite{dzi99}). This is in qualitative (but not quantitative !) agreement
with the observations.  The modulation period is not directly related to
the rotation period of the stars, but is determined by interplay between
the frequency spacing and rotational splitting of the nonradial modes
involved in the interaction.  As such, it can be different for the overtone
and for the fundamental-mode pulsators.

The outstanding difficulty of the resonant model is explaining the
existence of frequency doublets.  In some cases, the observed doublet can
actually be a triplet, with one sidepeak too small to be detected.  For
several stars, however, such an explanation would require an extreme
amplitude asymmetry in order to be valid. For example, we find $3\sigma$
upper limits of $A_{+}/A_{-} < 0.08$ for \object{BWC~V47} (RRc star) and
$A_{-}/A_{+} < 0.19$ for \object{BWC~V51} (RRab star).  It is not clear at present
if so strongly asymmetric triplets (or pure doublets) can be reproduced by
the resonant mode coupling theory.

\begin{acknowledgements}
This work has been supported in part by Polish KBN grants 2~P03D~002~20 and
5~P03D~012~20.
\end{acknowledgements}


\end{document}